\documentclass[12pt]{article}
\usepackage[english]{babel}
\usepackage[cp1251]{inputenc}
\usepackage[a4paper, left=25mm, right=15mm, top=20mm, bottom=20mm]{geometry}
\usepackage{graphicx}
\usepackage{amssymb}
\usepackage{xcolor}
\usepackage{hyperref}
\hypersetup{pdfstartview=FitH, linkcolor=linkcolor,citecolor=citecolor,colorlinks=true} 
\usepackage{amsmath}
\usepackage{setspace}
\usepackage{bm}
\usepackage{cite}
\usepackage{url}
\usepackage[hang,flushmargin]{footmisc} 
\newcommand{\be}{\begin{equation}}
\newcommand{\ee}{\end{equation}}
\renewcommand{\baselinestretch}{1.3}
\date{}
\begin{document}
\definecolor{linkcolor}{HTML}{008000}
\definecolor{citecolor}{HTML}{4682B4}
\large
\title{\bf \Large Hong-Ou-Mandel interference on a real beam splitter}
\normalsize
\author{Dmitry N. Makarov, \\
Northern (Arctic) Federal University, Arkhangelsk, 163002, Russia\\
E-mail: makarovd0608@yandex.ru  }

\maketitle
\begin{abstract}
\noindent Hong-Ou-Mandel (HOM) effect is known to be one of the main phenomena in quantum optics. The effect occurs when two identical single-photon waves enter a 1:1 beam splitter, one in each input port. When the photons are identical, they will extinguish each other. One of the main elements of the HOM interferometer is the beam splitter, which has its own coefficients of reflection $R = 1/2$ and transmission  $ T = 1/2 $. In this work, the general mechanism of the interaction of two photons in a beam splitter is considered using an analytical solution, which shows that in the HOM theory of the effect it is necessary to know (including when planning the experiment) not only $ R = 1/2 $ and $ T = 1/2 $, but also their root-mean-square fluctuations $ \Delta R ^ 2, \Delta T ^ 2 $, which arise due to the dependence of $R = R(\omega_1, \omega_2) $ and $ T = T (\omega_1, \omega_2) $ on the frequencies where $\omega_1, \omega_2$ are the frequencies of the first and second photons, respectively. Under certain conditions, specifically when the dependence of the fluctuations $ \Delta R^2 $ and $ \Delta T^2 $ can be neglected and $ R=T=1/2 $ is chosen, the developed theory coincides with previously known results.\\

\end{abstract}

\section{Introduction}
\begin{figure}[!h]
\center{\includegraphics[angle=0, width=0.7\textwidth, keepaspectratio]{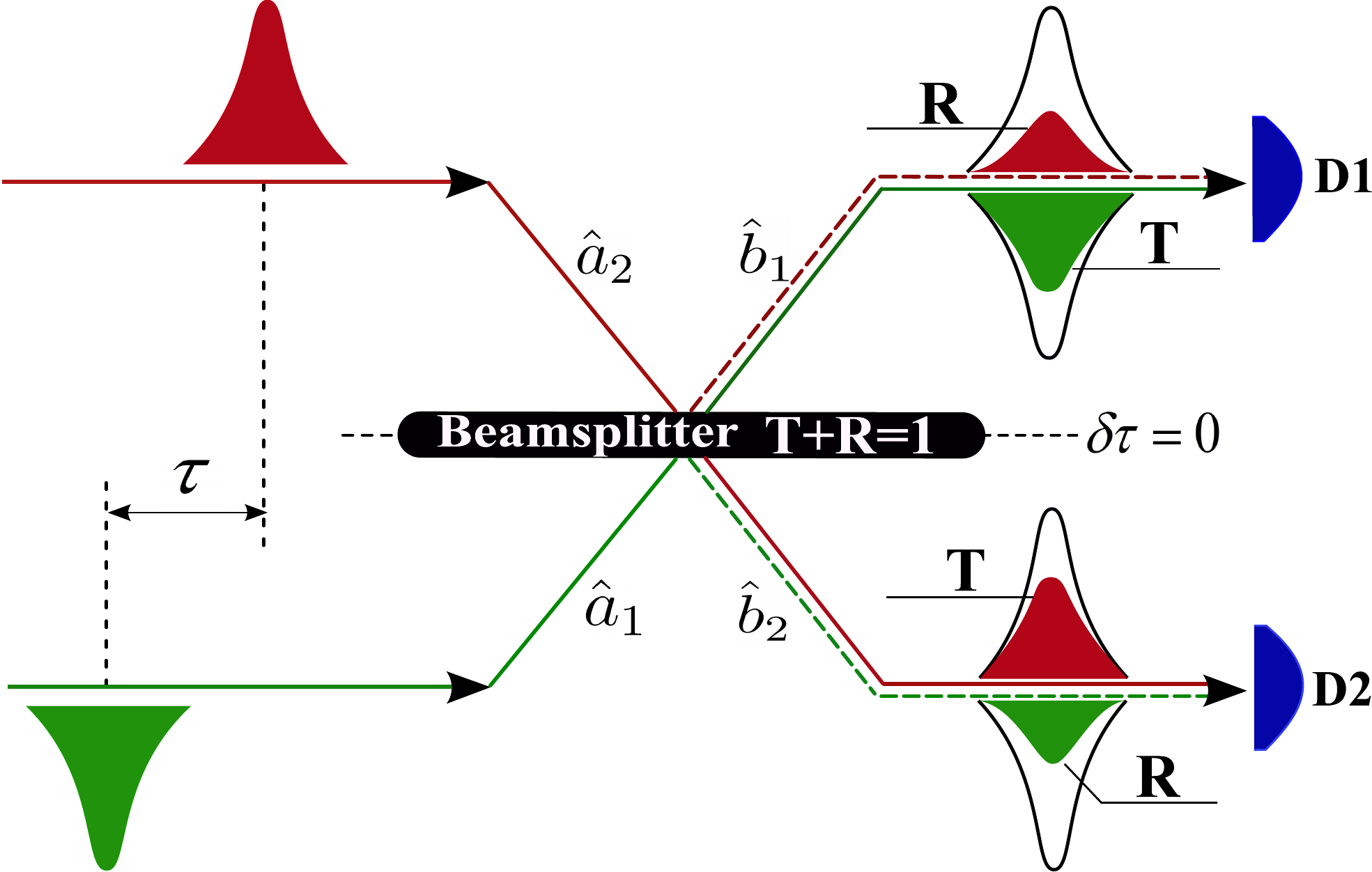}} 
\caption[fig_1]{Schematic representation of the HOM interferometer, where $D_1, D_2$ are the first and second detectors, respectively; $ \tau$ is the time delay between 1 and 2 photons and $ \delta \tau$ is the time delay caused by the spatial displacement of the BS from the equilibrium position.}
\label{fig_1}
\end{figure}

The HOM effect was first experimentally demonstrated by Hong et al in 1987 \cite{HOM_1987}. HOM interference shows up in many instances, both in fundamental studies of quantum mechanics and in practical implementations of quantum technologies \cite{Nielsen_2000,Gisin_2002,Sangouard_2011}. For example, one of the main practical applications of the HOM effect is to check the degree of indistinguishability of two incoming photons. When the HOM dip reaches all the way down to zero coincident counts, the incoming photons are perfectly indistinguishable, whereas if there is no dip, the photons are distinguishable.
A HOM interferometer scheme was presented in \cite{HOM_1987}, one of the main elements of which was a beam splitter (BS). To observe quantum interference, a beam splitter is chosen close to 1:1 (having coefficients of reflection $ R $ and transmission $ T $ close to 1/2).
A theoretical explanation of the HOM effect based on constant coefficients $ R $ and $ T $ and boson statistics of photons is quite simple \cite{Mandel_1995,Scully_1997}. In this interpretation, we are not interested in what happens to the incident photons in the beam splitter. For this, they consider BS lossless (hereinafter simply BS) as ideal, i.e. with constant coefficients $ R $ and $ T $ and BS is the source of the other two photons obeying bosonic statistics.  In this case, the annihilation operators before entering 1 and 2 photons in BS represent $ \hat{a}_1$ and $ \hat{a}_2$, respectively, and after exiting BS is $\hat{b}_1$ and $\hat{b}_2$. The transformation from one pair of operators to another is generally described by the BS matrix (denoted as $U_{BS}$) in the form (see, for example, \cite{Campos_1989,Luis_1995})
\begin{eqnarray}
\begin{pmatrix}
  \hat{b}_1\\
  \hat{b}_2
\end{pmatrix}=
U_{BS}
\begin{pmatrix}
  \hat{a}_1\\
  \hat{a}_2
\end{pmatrix}, ~~~ U_{BS}=\begin{pmatrix}
  e^{i\phi_1}\sqrt{T}&   e^{i\phi_2}\sqrt{R}\\
  -   e^{-i\phi_2}\sqrt{R}&    e^{-i\phi_1}\sqrt{T}
\end{pmatrix} .
\label{1}
\end{eqnarray}
It is easy to see that for $R=T=1/2$, the photons at the output (described by the operator $ \hat{b}_2\hat{b}_1$) only come out in pairs from 1 or 2 ports. This analysis is fundamental to understanding the HOM effect and is not subject to any additional research. The basic scheme HOM interferometer for arbitrary photons (including quantum entangled photons) is shown in Fig. \ref{fig_1}.
In reality, the pair of photons arriving at the BS do not have a set frequency, but have a certain frequency distribution. Nonetheless, in the theoretical description (for example, \cite{HOM_1987,Fearn_1989,Aephraim_1992,Legero_2003,Wang_2006,Legero_2004}) of the experimentally observed value $P$ ($P$ is the joint probability of detecting photons after exiting the BS on the output ports), the frequency distribution does not affect BS matrix $U_{BS}$, because $R$ and $T$ are constant values. Currently, the well-known HOM effect theories are based on calculating the value of $P$ within the constant values of $R=T=1/2$. However, some studies have been published on the dynamics of photon transport and photon interaction in various systems, for example, in two-level \cite{Longo_2010} in which the coefficients $R$ and $T$ are variables, but these studies are treated as not suitable for interpreting the HOM effect in terms of the photon interaction in BS.

In the work presented in this paper the coefficients $R$ and $T$ are variables, which significantly affects the theory of the HOM effect. The problem of interaction of two photons in BS is solved analytically, allowing the determination of the photon statistics after exiting the BS. Within the general form $U_{BS}$ is a BS matrix similar to (\ref{1}), where $R$ and $T$ are some functions that depend on the frequencies of incident photons, the interaction time of two photons in BS, and on the BS material. This leads to the value of $P$ being calculated to take into account the dependence on the frequencies of $R$ and $T$. It is shown that even in the case of identical incident photons and their average values $\bar{R}=\bar{T}=1/2$ (averaging over the frequencies of incident photons), a zero value of $P$ may not be observed, despite being predicted by the HOM interference theory taking into account the constants $R$ and $T$. Indeed, in the case of constant coefficients, as well as without a time delay between two photons, i.e. $ \delta \tau = 0$ or $\tau = 0$ and identical photons, because it is well known that $P \propto (R-T)^2$, for $R=T=1/2$ we get $P=0$ \cite{HOM_1987, Fearn_1989}. In our case, $P \propto \overline {(R-T)^2}$, which means $P \propto \overline{R^2}-(\overline{R})^2$ or $P \propto \overline{T^2}-(\overline{T})^2$ (when $\bar{R}=\bar{T}=1/2$) i.e. there is a fluctuation in the reflection and transmission coefficients, had was not earlier taken into account in theoretical and experimental studies.  It is shown (arbitrary falling photons, including not only Fock photons, but also taking into account the time delay $ \delta \tau$ and $\tau$) that under certain conditions the coefficients $R$ and $T$ can be considered constant, and the results obtained pass into well-known approaches. The theory developed here is especially important when planning experiments in the HOM interferometer and analyzing them; because the fluctuations of $R$ and $T$ can be very large, the results of such experiments may not be correctly interpreted.

\section{Photons in BS}
Consider a polyatomic system (for example, BS) interacting with two photons. We represent the electromagnetic field of photons through the transverse vector potential $ {\bf A} $ in the Coulomb gauge $ div {\bf A} = 0 $ \cite {Mandel_1995,Scully_1997}, then the Hamiltonian of such a system will be (further, the atomic system of units to be used will be: $\hbar$ = 1; $|e|$ = 1; $m_e$ = 1, where  $\hbar$ is Dirac’s constant, $e$ is the electron charge,  $m_e$ is the electron mass)
\begin{eqnarray}
\left\lbrace {\hat H}_{1}+{\hat H}_{2}+\frac{1}{2}\sum_a\left({\hat{\bf p}}_a+\frac{1}{c}\hat{\bf A}_a\right)^2 + \sum_a U({\bf r}_a)\right\rbrace \Psi = i\frac{\partial \Psi}{\partial t}, 
\label{2}
\end{eqnarray}
where $ {\hat H}_{i} =\omega_i\hat{a_i}^{+}\hat {a_i} $ is the Hamilton operator for the first ($ i=1 $) and the second ($ i = 2 $) photon ($ \omega_i $ is the frequency, and $ \hat{a_i} $ is the annihilation operator of the photon with number $ i $); $ U ({\bf r}_a) $ is the atomic potential acting on the electron with number $ a $ ; $ {\hat{\bf p}}_a $ is the electron momentum operator with the number $ a $; $ \hat{\bf A}_a = \hat{\bf A}_{1, a} + \hat{\bf A}_{2,a} $, where $ \hat {\bf A}_{i,a} = \sqrt{\frac{2 \pi c^2}{\omega_i V_i}}{\bf u}_i(\hat{a_i}^{+}+\hat{a_i}) $ is the vector potential in the dipole approximation created by the $ i $ photon acting on the electron with number $ a $ ($ c $ is the speed of light, $ V_i $ is the modal volume, $ {\bf u}_i $ is the polarization of the photon with the number th $ i $) \cite{Mandel_1995, Scully_1997}; and the sum $ \sum_a $ a (\ref{2}) is over all electrons polyatomic system. It should be added that the dipole approximation gives correct results at photon wavelengths $ \lambda \gg 1 $ i.e.  much larger than atomic sizes. Furthermore Eq. (\ref{2}) is more convenient to consider in the form of a differential equation (which was the approach taken in \cite{Makarov_2017_adf, Makarov_SREP_2018, Makarov_2019_PRA}), going from the operators $ {\hat a} = \frac {1}{\sqrt{2}}(q+\frac {\partial} {\partial q}), {\hat a}^{+} = \frac{1}{\sqrt {2}} (q-\frac{\partial}{\partial q}) $ to the electromagnetic field variables $ q $ \cite {Mandel_1995, Scully_1997}. As a result, the Hamiltonian of Eq. (\ref{2}) will be
\begin{eqnarray}
{\hat H}= \sum^{2}_{i=1}\left\lbrace \frac{\omega_i}{2}\left( q^2_i-\frac{\partial^2}{\partial q^2_i} \right)+\overline{N}\frac{\beta^2_i}{2}q^2_i+\beta_i q_i{\bf u}_i\sum_a {\hat{\bf p}_a}\right\rbrace+\overline{N}\beta_1 \beta_2 q_1 q_2 {\bf u}_1 {\bf u}_2+\sum_a\frac{{\hat{\bf p}}^2_a}{2}+\sum_a U({\bf r}_a), 
\label{3}
\end{eqnarray}
where $ \beta_i = \sqrt{\frac{4 \pi}{\omega_i V_i}} $, and the value $ \overline{N}= \sum_a (1) $ is the number of electrons participating in the interaction with photons in a polyatomic system.
Eq. (\ref{3}) can be seen to correspond to the equation for coupled harmonic oscillators interacting with the electrons of a polyatomic system. A similar system was considered in \cite{Makarov_2018_PRE}, but without taking into account interaction with electrons, we obtain 
\begin{eqnarray}
{\hat H}= \sum^{2}_{i=1} \frac{1}{2}\left\lbrace \sqrt{A_i} \left( {\hat P}^2_i+y^2_i\right)  +2 {\bf D}_i y_i\sum_a {\hat{\bf p}_a}\right\rbrace + \sum_a \frac{{\hat{\bf p}}^2_a}{2}+\sum_a U({\bf r}_a), 
\label{4}
\end{eqnarray}
where $ y_1 =  A^{1/4}_1\left( q_1/\sqrt{B_1} \cos \alpha-q_2/\sqrt{B_2} \sin \alpha\right)$ and $y_2 = A^{1/4}_2\left(q_1/\sqrt{B_1} \sin \alpha + q_2/\sqrt{B_2} \cos \alpha \right)$ is new variables; ${\hat P}_i=-i\partial/\partial y_i$;  ${\bf D}_1=\beta_1 A^{-1/4}_1\sqrt{B_1}{\bf u}_1\cos\alpha - \beta_2  A^{-1/4}_1\sqrt{B_2}{\bf u}_2 \sin\alpha$; ${\bf D}_2=\beta_1 A^{-1/4}_2\sqrt{B_1}{\bf u}_1\sin\alpha+\beta_2A^{-1/4}_2 \sqrt{B_2}{\bf u}_2 \cos\alpha$. The study \cite{Makarov_2018_PRE} showed that $\tan(2\alpha)=C/(B^2_2-B^2_1)$, where $C=2\overline{N} \beta_1 \beta_2 \sqrt{B_1 B_2} {\bf u}_1 {\bf u}_2$; $B_i=\omega_i+\overline{N}\beta^2_i$ and $A_i=B^2_i+(-1)^i C/2 \tan\alpha$.
Obviously, the value of $ \beta_i $ is very small in the case of single-photon interaction, see, for example, \cite{Tey_2008}, where $ \beta \ll 1 $, even in the case of strong focusing. In this case, the quantities $ {\bf D}_1 $ and $ {\bf D}_2 $ are negligible. This is an obvious fact, since these quantities are responsible for various inelastic transitions of electrons in an atom under the action of photons, which are usually negligible in lossless BS. As a result, the dynamics of two photons in BS will be described by the wave function
\begin{eqnarray}
|\Phi(t_{BS})\rangle =e^{-i {\hat H_{BS}}t_{BS}}|\Phi(0)\rangle  , ~~~ \hat H_{BS} = \sum^{2}_{i=1} \frac{\sqrt{A_i}}{2}\left\lbrace {\hat P}^2_i+y^2_i\right\rbrace ,
\label{5}
\end{eqnarray}
where $ t_{BS} $ is the photon interaction time in BS and $ |\Phi(0)\rangle $ is the initial state of the photons before entering the BS. It is noteworthy that for small $ \beta_i $ the parameter $ \alpha $ can have non small values at close photon frequencies (or identical), i.e. their interaction is significant. If the frequencies differ by a value much greater than $ \beta^2 $, the parameter $ \alpha \to 0 $ and there is no photon interaction in BS, i.e. no quantum interference occurs.

In the future, to calculate the required quantities, we will need the eclectic field operators $ {\hat E}^{+}_1(t_1) $ and $ {\hat E}^{+}_2(t_2) $ at time instants $ t_1 $ and $ t_2 $ on the first and second detectors, respectively. To do this, we need to find the evolution (in BS, as well as from BS to detectors) of the operators $ {\hat E}^{+}_{01}(0) $ and ${\hat E}^{+}_{02}(0) $ of the first and second photons, respectively
\begin{eqnarray}
{\hat E}^{+}_1(t_1)=e^{i {\hat H_0}t_1}e^{i {\hat H_{BS}}t_{BS}}{\hat E}^{+}_{01}(0)e^{-i {\hat H_{BS}}t_{BS}} e^{-i {\hat H_0}t_1},
\nonumber\\
{\hat E}^{+}_2(t_2)=e^{i {\hat H_0}t_2}e^{i {\hat H_{BS}}t_{BS}}{\hat E}^{+}_{02}(0)e^{-i {\hat H_{BS}}t_{BS}} e^{-i {\hat H_0}t_2},
\label{6}
\end{eqnarray}
where $ {\hat H_0} = \sum^{2}_{i = 1} \omega_i/2\left \lbrace - \partial^2/ \partial q^2_i + q^2_i \right \rbrace $ is the Hamiltonian of photons outside of BS. Because $ {\hat E}^{+}_{01}(0) \propto{\hat a}_1 $, and $ {\hat E}^{+}_{02}(0) \propto {\hat a}_2 $ (see, for example, \cite{Mandel_1995,Scully_1997}), it is more convenient to consider not the eclectic field operators, but the photon creation and annihilation operators before entering BS ($ {\hat a}_1 $ and $ {\hat a}_2 $) and on the detectors ($ {\hat b}_1 $ and $ {\hat b}_2 $). To this end, we replace $ {\hat E}^{+}_{01}(0)\to{\hat a}_1, {\hat E}^{+}_{02}(0)\to{\hat a}_2 $ and $ {\hat E}^{+}_{1} (t_1) \to{\hat b}_1(t_1), {\hat E}^{+}_{2}(t_2) \to{\hat b}_2(t_2) $. Taking into account the time delay $ \delta \tau $ for the spatial displacement of BS from the equilibrium position and the time delay $ \tau $ between 1 and 2 photons (see Fig. \ref{fig_1}) (see Appendix)
\begin{eqnarray}
{\hat b}_1(t_1)=e^{i\phi_1}\sqrt{T}  e^{-i\omega_1 (t_1-\tau)}{\hat a}_1+e^{i\phi_2}\sqrt{R} e^{-i\omega_2 (t_1+\delta \tau/2)}{\hat a}_2,
\nonumber\\
{\hat b}_2(t_2)=e^{-i\phi_1}\sqrt{T} e^{-i\omega_2 t_2}{\hat a}_2 -e^{-i\phi_2}\sqrt{R} e^{-i\omega_1 (t_2-\delta \tau/2-\tau)}{\hat a}_1,
\label{7}
\end{eqnarray}
where $ \phi_1, \phi_1 $ are some non-essential phases, and the coefficients
\begin{eqnarray}
T=\frac{1+2\epsilon^2+\cos\left(\Omega t_{BS} \sqrt{1+\epsilon^2} \right) }{2(1+\epsilon^2)},~~R=\frac{1-\cos\left( \Omega t_{BS} \sqrt{1+\epsilon^2}  \right) }{2(1+\epsilon^2)},
\nonumber\\
\Omega=\frac{8\pi \overline{N} {\bf u}_1 {\bf u}_2 \sqrt{B_1 B_2} }{(B_1+B_2)\sqrt{V_1 V_2 \omega_1 \omega_2}},~~ \epsilon=\frac{B_2 -B_1}{\Omega}.~~~~~~~~~~~~~~~~~~
\label{8}
\end{eqnarray}
From (\ref{7}) it can be seen that the matrix BS that is $ U_ {BS} $ completely corresponds to the matrix (\ref{1}) (needless to say, for $ t_1 = t_2 = t $ and for $ \delta \tau = \tau = 0 $). It should also be added that the coefficients $ T $ and $ R $ now depend on the frequencies, but the condition $ R + T = 1 $ is still satisfied. In addition, $ T $ and $ R $ are symmetric, i.e. if we change the first to the second photon $ \omega_1 \to \omega_2, V_1 \to V_2, {\bf u}_1 \to{\bf u} _2 $ and vice versa, then, as anticipated, the coefficients will not alter.

\section{Two-photon interference}
The next problem, we consider is the probability $ P_ {1,2} $ of the joint detection of photons on 1 and 2 detectors (correlation between the two detectors). If our coincidence gate window accepts counts for a time $ T_D $, then the rate of coincidences P, between detectors 1 and 2 is proportional to (see, for example, \cite{HOM_1987, Fearn_1989, Aephraim_1992})
\begin{eqnarray}
P_{1,2}\propto \int^{T_{D}/2}_{-T_{D}/2}\int^{T_{D}/2}_{-T_{D}/2}\langle {\hat b}^{\dagger}_1(t_1){\hat b}^{\dagger}_2(t_2){\hat b}_1(t_1){\hat b}_2(t_2) \rangle dt_1 dt_2.
\label{9}
\end{eqnarray}
Let us consider the case where the reaction time $ \tau_D $ (time resolution) of the detectors $ D_1 $ and $ D_2 $ in the experiment is many times slower than other time scales of the problem $ \tau_D \gg 1 $: in this case $ T_D \to \infty $. It should be added that the theory presented below is not difficult to generalize to the case of $ \tau_D \ll 1 $, which is currently implemented experimentally (for example, \cite{Legero_2004, Lyons_2018}.

Eq. (\ref {9}) is applicable in the case of monochromatic photons. In reality, they cannot be such and it is necessary to take into account the frequency distribution, and in this case the initial wave function of the photons will be in the form $ | \Psi \rangle = \int \phi(\omega_1, \omega_2) {\hat a}^{\dagger}_2{\hat a}^{\dagger}_1 | 0 \rangle d \omega_1 d \omega_2 $, where $ \phi(\omega_1,\omega_2) $ is the joint spectral amplitude (JSA) of the two-photon wavefunction ($ \int | \phi (\omega_1, \omega_2) |^2 d \omega_1 d \omega_2 = 1 $). Further calculations of $ P_{1,2} $ are similar to those that are generally accepted (for example, \cite{Fearn_1989, Legero_2003, Aephraim_1992}), the only difference being that it is necessary to consider the $ T $ and $ R $ functions depending on the frequencies . As a result, we obtain
\begin{eqnarray}
P_{1,2}=\int^{\infty}_{-\infty}\int^{\infty}_{-\infty} \left|\xi_1(t_1,t_2,\tau)-\xi_2(t_1,t_2,\tau,\delta \tau) \right|^2  dt_1 dt_2~,
\nonumber\\
\xi_1(t_1,t_2,\tau)=\frac{1}{2\pi} \int \phi(\omega_1,\omega_2) e^{-i\omega_1(t_1-\tau)}e^{-i\omega_2 t_2} T(\omega_1,\omega_2) d\omega_1 d \omega_2~,
\nonumber\\
\xi_2(t_1,t_2,\tau,\delta \tau)=\frac{1}{2\pi} \int \phi(\omega_1,\omega_2) e^{-i\omega_2(t_1+\delta \tau)}e^{-i\omega_1(t_2-\delta \tau -\tau)} R(\omega_1,\omega_2) d\omega_1 d \omega_2 ~,
\label{10}
\end{eqnarray}
where $ P_{1,2} $ is normalized so that with $ t_{BS} = 0 $ the probability is $ P_{1,2} = 1 $ (without BS, the probability of joint operation of the detectors is $ 100 \% $), which corresponds to standard normalization in HOM theory. We then obtain
\begin{eqnarray}
P_{1,2}=\int \Biggr(  |\phi(\omega_1,\omega_2)|^2 \left( T^2(\omega_1,\omega_2)+R^2(\omega_1,\omega_2) \right) - 
\nonumber\\
-2{\rm Re}\biggr\lbrace \phi(\omega_1,\omega_2)\phi^*(\omega_2,\omega_1) T(\omega_1,\omega_2)R(\omega_2,\omega_1)e^{-i(\omega_2-\omega_1)(\delta \tau +\tau)}\biggl\rbrace \Biggl) d \omega_1 d \omega_2  .
\label{11}
\end{eqnarray}
It should be added that if $ T $ and $ R $ are assumed to be independent of frequencies and $ T = R = 1/2 $, then Eq. (\ref{11}) corresponds to the well-known equation, for example, \cite{Grice_1997,Erdmann_2000 ,Barbieri_2017}. It is also seen that the time delay of $ \delta \tau $ and $ \tau $ is additively $ \delta \tau + \tau $; therefore, we denote it by $ \Delta \tau = \delta \tau + \tau $.

We next consider the case of identical photons at $ \Delta \tau = 0 $, in this case $ \phi (\omega_1, \omega_2) = \phi(\omega_2, \omega_1) $ and $ R(\omega_1, \omega_2) = R (\omega_2, \omega_1) $ (because $ V_1 = V_2 $), and the quantity
\begin{eqnarray}
P_{1,2}(\Delta \tau=0)=\int   |\phi(\omega_1,\omega_2)|^2 \left( T(\omega_1,\omega_2)-R(\omega_1,\omega_2) \right)^2  d \omega_1 d \omega_2 =\overline{(T-R)^2} .
\label{12}
\end{eqnarray}
If in (\ref{12}) we choose $ \overline{T} = \overline{R} = 1/2 $, then we get $ P_{1,2} = 4 (\overline {T^2} - \overline{T}^2) = 4 (\overline{R^2} - \overline{R}^2) $. In other words, there is a mean-square fluctuation of the coefficients of transmission $ T $ and reflection $ R $, which leads to a nonzero value of $ P_ {1,2} $ in the case of identical photons. This conclusion is fundamental in the theory of HOM interference and was not previously obtained, because it was believed that if the photons are identical, $ T = R = 1/2 $ and $ \Delta \tau = 0 $, then $ P_{1,2} = 0 $. Also, from the previously obtained Eqs. (\ref{11}) and (\ref{12}) it follows that the $ P_{1,2}(\Delta \tau \gg \tau_c)=2\overline{T^2}=2\overline{R^2}$ ($ \tau_c $ is the coherence time), as well as $P_{1,2}(\Delta \tau \gg \tau_c)=1/2(1+P_{1,2}(\Delta \tau=0))$.

\begin{figure}[h]
\center{\includegraphics[angle=0,width=1\textwidth, keepaspectratio]{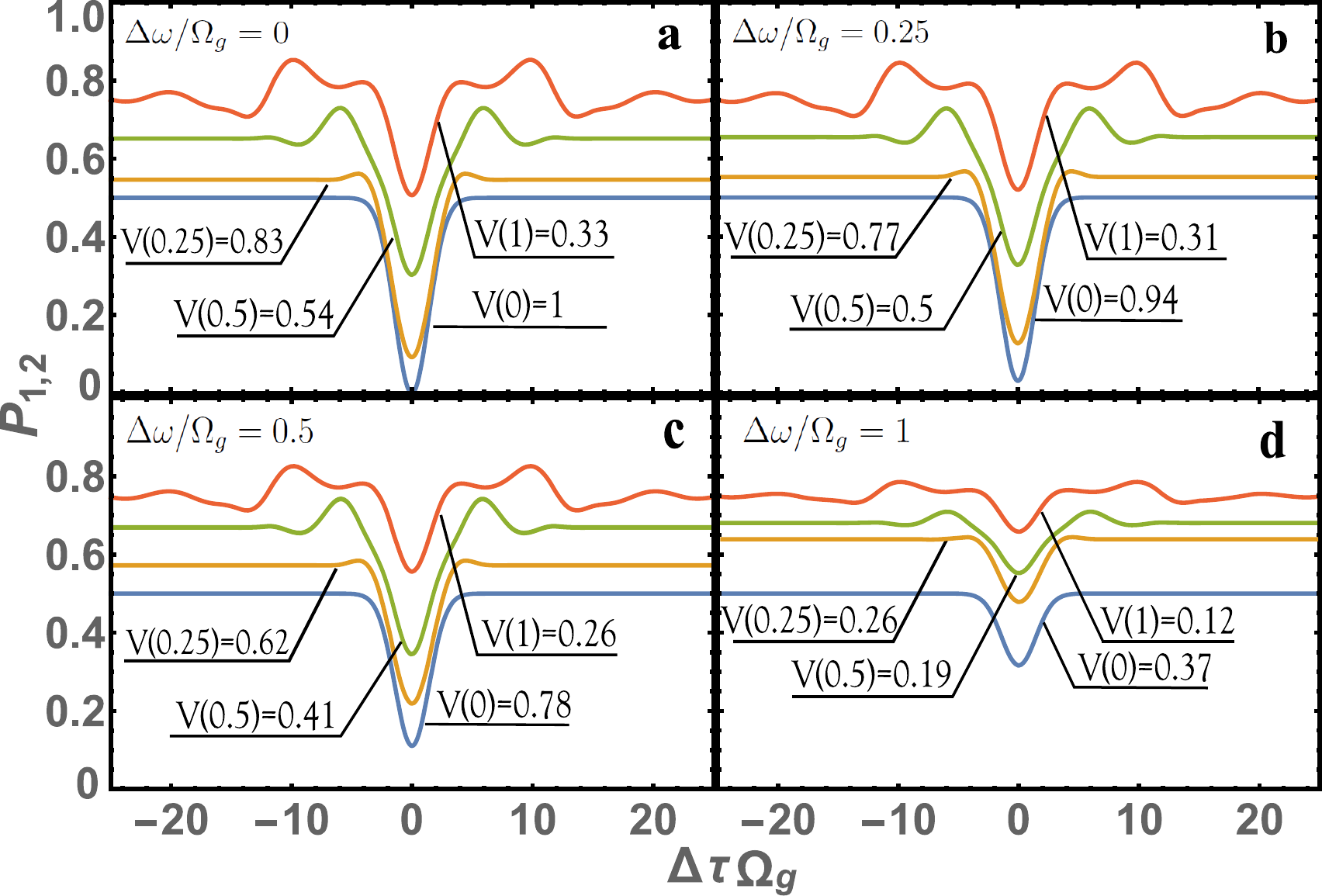}} \\
\caption[fig_2]{Dependence of $ P_{1,2} $ on $ \Delta \tau \Omega_g $ (HOM dip). Case (a) corresponds to completely identical photons, and cases (b), (c), (d) correspond to non-identical photons. The visibility $ {\rm V} = {\rm V} (\Omega_g / \Omega) $ depends on the parameter $ \Omega_g / \Omega $. The case $ \Omega_g / \Omega = 0 $ and visibility $ {\rm V} (0) $ corresponds to the previously known theory of HOM interference with constant coefficients $ T = R = 1/2 $.}
\label{fig_2}
\end{figure}

Let us present the results of calculating the value of $ P_{1,2} $ for the case
\begin{eqnarray}
\phi(\omega_1,\omega_2)=\frac{1}{\sqrt{\pi}}\frac{(\sigma^2_1+\sigma^2_2+\sigma^2_p)^{1/4}}{\sqrt{\sigma_1 \sigma_2 \sigma_p}}e^{-\frac{(\omega_1+\omega_2-\Omega_p)^2}{2\sigma^2_p}}e^{-\frac{(\omega_1-\omega_{01})^2}{2\sigma^2_1}}e^{-\frac{(\omega_2-\omega_{02})^2}{2\sigma^2_2}} .
\label{13}
\end{eqnarray}
The function (\ref {13}) allows us to analyze the value of $ P_{1,2} $ for two cases that are of practical interest. The first case is SPDC, for example, for $ \Omega_p = 2 \omega_0, \omega_0 = \omega_{01} = \omega_{02}, \sigma_1 = \sigma_2 = \sigma $ is SPDC of type I, where $ \sigma_p $ is the bandwidth of the pump beam, $ \omega_0 $ and $ \sigma $ are the central frequency and the bandwidth, respectively, for both the signal and the idle beams \cite{Shih_1999}. If we consider $ \sigma_p \to \infty $ in (\ref{13}), then this will be the case of Fock photons. Substituting (\ref{13}) into (\ref{11}) we obtain
\begin{eqnarray}
P_{1,2}= \int^{\infty}_{-\infty}\biggr\lbrace e^{-(y-\frac{\Delta \omega}{\Omega_g})^2}  \left(  T^2(y)+R^2(y)\right) -
2 B e^{-(\frac{\Delta \omega}{\Omega_g})^2}T(B y)R(-B y) e^{-y^2}\cos \left(B \Delta \tau \Omega_g y \right) \biggr\rbrace \frac{d y}{\sqrt{\pi}} ,
\nonumber\\
B=A\sqrt{\frac{1+\frac{\sigma^2_p}{\sigma^2_1+\sigma^2_2}}{A^2+\frac{\sigma^2_p}{\sigma^2_1+\sigma^2_2}}},~A=\frac{2\sigma_1\sigma_2}{\sigma^2_1+\sigma^2_2},~\Delta \omega=\omega_{02}-\omega_{01},~\Omega_g=\sqrt{\frac{4\sigma^2_1\sigma^2_2+(\sigma^2_1+\sigma^2_2)\sigma^2_p}{\sigma^2_1+\sigma^2_2+\sigma^2_p}},~~~~~~~~~~~
\label{14}
\end{eqnarray}
where $ B \in (0,1) $, and $ T (y) $ and $ R (y) $ are determined by the Eq. (\ref{8}), with the only difference being that
\begin{eqnarray}
\Omega=\frac{4\pi \overline{N} {\bf u}_1 {\bf u}_2 }{\omega_0\sqrt{V_1 V_2}}\frac{\sqrt{1+\frac{4\pi \overline{N}}{V_1\omega^2_0}} \sqrt{1+\frac{4\pi \overline{N}}{V_2\omega^2_0}}}{1+\frac{2\pi \overline{N}}{\omega^2_0}\left(\frac{1}{V_1}+\frac{1}{V_2}\right)},~\omega_0=\frac{\omega_{01}+\omega_{02}}{2},~ \epsilon=\frac{\Omega_g}{\Omega}y+\frac{4\pi \overline{N}}{\omega_0 \Omega}\left(\frac{1}{V_2}-\frac{1}{V_1}\right).
\label{15}
\end{eqnarray}
If we assume that $ \Omega_g / \Omega \ll 1 $, then $ T $ and $ R $ become constant values and they can always be selected in the experiment $ T = R = 1/2 $. The equation for $ P_{1,2} $, in our case Eq. (\ref{14}) for the constants $ T = R = 1/2 $ easily integrates and coincides with the well-known $ P_{1,2} = 1/2 ( 1-B e^{-(\Delta \omega / \Omega_g)^2} e^{- 1/4 (B \Omega_g \Delta \tau) ^ 2}) $, for example, \cite{Wang_2006}. Next, we consider how the value of $ P_{1,2} $ will look like depending on $ \Delta \tau \Omega_g $ (HOM dip) in the case of $ V_1 = V_2, \sigma_1 = \sigma_2 $ for different values of $ \Omega_g / \Omega $ and $ \Delta \omega / \Omega_g $, but for $ \Omega t_{BS} $ such that $ \overline{T} = \overline{R} = 1/2 $, see Fig. \ref{fig_2}: as $ \Omega_g / \Omega $ increases, the value of $ P_{1,2} $ tends to unity.
This can be seen in the general analysis of the Eqs. (\ref{8}) and (\ref{11}). Fig. \ref{fig_2} also shows that when $ T $ and $ R $ are taken into account from the frequency, $ P_ {1,2} $ can significantly differ from the previously known theory of HOM interference.
It should be added that in experiments most often there is no good coincidence of $ P_{1,2} = P_{1,2}(\Delta \tau) $ with theoretical predictions of HOM interference with constant coefficients $ T = R = 1/2 $ . In such experimental studies, additional oscillations of the dependence $ P_{1,2} = P_{1,2} (\Delta \tau) $ between the minimum of this function and $ P_{1,2} = P_{1,2} (\Delta \tau \gg \tau_c) $ ($ \tau_c $ is the coherence time). The presented theory perfectly explains this experimental fact (see, for example, \cite{Kwiat_1992, Tambasco1_2018}), which is usually not investigated and does not have a clear explanation.
\section{Discussion and Conclusion}
Thus, the developed theory shows that for a real BS, the coefficients of transmission $ T $ and the refraction of $ R $ depend on the frequency. This dependence can significantly change the well-known theory of HOM interference. Under certain conditions, when the dependence on frequencies can be neglected (for example, in the case (\ref{13}) for $ \Omega_g / \Omega \ll 1 $), the coefficients $ T = R = 1/2 $ can be selected, and the developed theory is the same as that applying to the case of an ideal BS.
In the special case of mixed, identical, and separable photons, there is a  relationship between the visibility ${\rm V}$ of the HOM dip and the purity $\mathcal{P}$ of the input photons when $T=R=1/2$  \cite{Scully_1997,Kwiat_1992,Law_2000,Ekert_2002,Lee_2003}
\begin{eqnarray}
{\rm V}=\frac{P_{1,2}(\Delta \tau \gg \tau_c)-P_{1,2}(\Delta \tau=0)}{P_{1,2}(\Delta \tau \gg \tau_c)}={\rm Tr}  \rho_1  \rho_2,~~{\rm V}( \rho_1= \rho_2)=\mathcal{P},
\label{16}
\end{eqnarray}
where $ \rho_1, \rho_2 $ are the density matrices of some quantum states of 1 and 2 photons, respectively.
If we take into account the dependence of $ T $ and $ R $ on frequency, it is easy to see that the dependences of $ {\rm V} = {\rm Tr} \rho_1 \rho_2 $ in (\ref {16}) will no longer apply (see Eq. (\ref {11}), where $ T $ and $ R $ are present). This leads to the important conclusion that visibility $ {\rm V} $ with significant fluctuations of the coefficients $ T $ and $ R $ cannot be used to judge quantum interference, and for $ \rho_1 = \rho_2 $ purity $ \mathcal {P} $ of the input photons. Therefore, when conducting an experiment, it is necessary to not only choose $ \overline {T} = \overline {R} = 1/2 $, but also minimize fluctuations. In addition, if fluctuations cannot be minimized, then the theoretical curve can be adjusted to that obtained experimentally, by using Eqs. (\ref {11}) and (\ref {8}), while finding $ \phi (\omega_1, \omega_2) $. Knowing $ \phi (\omega_1, \omega_2) $, you can find $ {\rm V} $ using the well-known equation $ {\rm V} = {\rm Re} \int \phi (\omega_1, \omega_2) \phi^* (\omega_2, \omega_1) d \omega_1 d \omega_2 $ for an ideal BS. It should be added that fluctuations in the HOM interference had not been previously measured, because it was believed that the beam divider had strictly specified coefficients $ T $ and $ R $ during the experiment. In the case of HOM interference, the coefficients $ T = R = 1/2 $ were selected, which actually correspond to $ \overline {T} = \overline {R} = 1/2 $ in the experiment. It is quite simple to measure fluctuations at $ \overline{T} = \overline{R} = 1/2 $, for this it is necessary to measure $ P_{1,2} $ at $ \Delta \tau \gg \tau_c $, because, $ P_ {1,2} = 2 \overline {T^2} = 2 \overline{R^2} $ (this can be seen from the equation (\ref{11}) for $ R + T = 1 $). If the fluctuations are small, then $ P_ {1,2} $ will go over to the known value $ P_{1,2} (\Delta \tau \gg \tau_c) = 1/2 $ (half of the maximum possible). It is also easy to find visibility $ {\rm V} $ in the case of identical photons at $ T = R = 1/2 $ using (\ref {16}) and $P_{1,2}(\Delta \tau \gg \tau_c)=2\overline{T^2}=2\overline{R^2}$ , $P_{1,2}(\Delta \tau=0)=4(\overline{T^2}-\overline{T}^2)=4(\overline{R^2}-\overline{R}^2)$  as ${\rm V}_{id}=2\overline{R}^2/\overline{R^2}-1=2\overline{T}^2/ \overline{T^2}-1$. 

A new physical quantity appears in the theory presented, which characterizes BS and its interaction with photons (\ref{8}) is $ \Omega $. This value depends on the characteristics of the incident photons: $ \omega_1, \omega_2, V_1, V_2, {\bf u}_1, {\bf u}_2 $, and on the characteristics of BS itself  $ \overline{N} $. This value is the frequency response of the photon interaction in BS. Three cases of photon interaction in BS can be distinguished. The first case is when the frequency (mean the characteristic $ B_i \sim \omega_i $) difference between the incident photons is many times less than $ \Omega $ i.e. $ B_2- B_1 \ll \Omega $, see (\ref{8}); then the interaction of photons in BS occurs so that the coefficients $ T $ and $ R $ are frequency independent. The second case is when the frequency difference between the incident photons is comparable to $ \Omega $ i.e. $ B_2- B_1 \sim \Omega $; then the interaction of photons in BS occurs so that the coefficients $ T $ and $ R $ depend on the frequencies. The third case is when the frequency difference between the incident photons is many times greater than $ \Omega $ i.e. $ B_2- B_1 \gg \Omega $; then in this case the interaction of photons in the BS does not occur $ R \to 0, T \to 1 $, in other words, in this case the BS is transparent to photons. The frequency $ \Omega $ can be estimated if we consider a pair of photons that is quite close in characteristics, i.e. $ V_1 \approx V_2 \approx V $, as well as $ \omega_1 \approx \omega_2 \approx \omega_0, {\bf u} _1 {\bf u} _2 \approx 1 $ and assume that the overlap of the photon wave packets in BS is ideal (all electrons of atoms with the number $ \overline{N} $ are in the volume $ V $). With such an estimate, it is easy to obtain that $ \Omega_{id} = 4 \pi n / \omega_0$, where $ n = \overline{N}/V $ is the electron concentration in BS ($ \Omega_ {id} $ is $ \Omega $ in the case of identical photons). If we quantify $ \Omega_ {id} $ for solid materials and the optical frequency range, we get that $ \Omega_ {id} \sim (10^{14} - 10^{17}) {\rm rad/s} $. Obviously, a similar estimate is also valid for non-identical photons; the order in such an estimate will be preserved i.e. $ \Omega \sim \Omega_ {id} $. In reality, these $ \Omega $ values have lower values due to non-ideal overlap of the wave packets of photons in the BS. It can be seen that these values of $ \Omega $ are essential in the theory of HOM interference. For example, from the Eqs. (\ref{14}) and (\ref{15}) it can be seen that the dependence of $ T $ and $ R $ on frequencies is determined by the relation $ \Omega_s / \Omega $, considering the case of optical frequencies of photons with $ \omega_0 \sim 10^{15} {\rm rad/s} $, where $ \Omega_s $ is usually less by orders of magnitude than $ \omega_0 $ (for example, in \cite{HOM_1987}, the $\Omega_s \approx 2\pi/\tau_c\approx 10^{14} {\rm rad/s} $ value was obtained) we get what could be $ \Omega_s / \Omega \sim 1 $ (this is the second case with $ \omega_2- \omega_1 \sim \Omega $). There is also an experimental possibility of determining the real frequency of $ \Omega_ {id} $, which means knowing the order of $ \Omega $. For this, it is necessary to use photons close to identical with the known spectral amplitude characteristics $ \phi(\omega_1, \omega_2) $ and measure $ P_{1,2} (\Delta \tau \gg \tau_c) = 2 \overline {T^2} = 2 \overline{R^2} $ for $ \overline {T} = \overline {R} = 1/2 $. Knowing this data, it is easy to find not only $ \Omega_ {id} $ using (\ref{8}), but also $ t_ {BS} $. You can also find these characteristics by measuring instead of $ P_ {1,2} (\Delta \tau \gg \tau_c) $ the value of $ P_{1,2} $, but with $ \Delta \tau = 0 $, i.e. $ P_ {1,2} (0) $. It can be seen from $\Omega_ {id}\sim 1/\omega_0$ that the developed theory is especially relevant in the case of the ultraviolet and X-ray frequency ranges, since $ \Omega_ {id} $ becomes smaller.


\begin{spacing}{0.5}

\end{spacing}

\newpage
\section*{Appendix}
\def\theequation{Ap.\arabic{equation}}
\setcounter{equation}{0}

Consider ${\hat E}^{+}_1(t_1)$ and ${\hat E}^{+}_2(t_2)$ in Eqs. (\ref{6}) articles (i.e. we will search for $ {\hat b}_1 (t_1)$ and $ {\hat b}_2 (t_2)$).  In this equation, we need to find ${\hat b}_i (t_i)=e^{{\rm i} {\hat H_0}t_i}e^{{\rm i} {\hat H_{BS}}t_{BS}}{\hat a}_{i}e^{-{\rm i} {\hat H_{BS}}}e^{-{\rm i} {\hat H_0} t_i} $ where ${\hat a}_{i}=1/\sqrt{2} (q_i+\partial/\partial q_i)$. Next, the operators ${\hat a}_{1}$ and ${\hat a}_{2}$ must be represented in the new variables $y_1,y_2$ (see Eq. (\ref{4}) and below).To observe quantum interference (HOM interference), we must consider such photon characteristics that the condition $B_2-B_1 \ll (B_1, B_2)$ is fulfilled. In this case, the value $ \tan(2\alpha)$ (defined below the Eq. (\ref{4})) takes non-zero values for $C \sim B^2_2-B^2_1$. This leads to $B_2-B_1 \ll (B_1, B_2)$ simplifying the variables $y_1, y_2$ as $y_1=q_1\cos\alpha-q_2\sin\alpha; y_2=q_1\sin\alpha+q_2\cos\alpha $. As a result using the known commutation rules it is easy to get
\begin{eqnarray}
e^{{\rm i} {\hat H_{BS}}t_{BS}}{\hat a}_{1}e^{-{\rm i} {\hat H_{BS}}t_{BS}}={\hat b}_{1}(0)=\cos\alpha e^{-{\rm i}\sqrt{A_1} t_{BS}}{\hat a}^{'}_1 +\sin\alpha e^{-{\rm i}\sqrt{A_2} t_{BS}}{\hat a}^{'}_2~,
\nonumber\\
e^{{\rm i} {\hat H_{BS}}t_{BS}}{\hat a}_{2}e^{-{\rm i}{\hat H_{BS}}t_{BS}}={\hat b}_{2}(0)=\cos\alpha e^{-{\rm i}\sqrt{A_2} t_{BS}}{\hat a}^{'}_2 -\sin\alpha e^{-{\rm i}\sqrt{A_1} t_{BS}}{\hat a}^{'}_1~,
\label{S1}
\end{eqnarray}
where ${\hat a}^{'}_{i}=1/\sqrt{2}(y_i+\partial/\partial y_i)$. Next, going back to the variables $q_1,q_2$ and using the known commutation rules, we get (discarding the non-essential phase)
\begin{eqnarray}
{\hat b}_{1}(t_1)=e^{-{\rm i}\omega_1 t_{1}}(\cos^2\alpha+\sin^2\alpha e^{-{\rm i}\Delta \omega t_{BS}}){\hat a}_1 + e^{-{\rm i}\omega_2 t_{1}}\sin\alpha \cos\alpha (1-e^{{\rm i}\Delta \omega t_{BS}}){\hat a}_2~,
\nonumber\\
{\hat b}_{2}(t_2)=e^{-{\rm i}\omega_2 t_{2}}(\cos^2\alpha+\sin^2\alpha e^{{\rm i}\Delta \omega t_{BS}}){\hat a}_2 - e^{-{\rm i}\omega_1 t_{2}}\sin\alpha \cos\alpha (1-e^{-{\rm i}\Delta \omega t_{BS}}){\hat a}_1~,
\label{S2}
\end{eqnarray}
where $ \Delta \omega = \sqrt{A_2}-\sqrt{A_1}$. The expression for $ \Delta \omega$ can be simplified by representing $ \Delta \omega =(A_2-A_1)/(\sqrt{A_2}+\sqrt{A_1}) $ and taking into account the above conditions $B_2-B_1 \ll (B_1, B_2)$ and $C\sim B^2_2-B^2_1$.   As a result, the Eq. (\ref{S2}) can be represented as
\begin{eqnarray}
{\hat b}_{1}(t_1)=e^{{\rm i}\phi_1}\sqrt{T}e^{-{\rm i}\omega_1 t_{1}}{\hat a}_1 + e^{{\rm i}\phi_2}\sqrt{R}e^{-{\rm i}\omega_2 t_{1}}{\hat a}_2~,~~~~~~~~~~~~~~~
\nonumber\\
{\hat b}_{2}(t_2)=e^{-{\rm i}\phi_1}\sqrt{T}e^{-{\rm i}\omega_2 t_{2}}{\hat a}_2 - e^{-{\rm i}\phi_2}\sqrt{R}e^{-{\rm i}\omega_1 t_{2}}{\hat a}_1~,~~~~~~~~~~~~
\nonumber\\
\phi_1={\rm Arg} (\cos^2\alpha+\sin^2\alpha e^{-{\rm i}\Delta \omega t_{BS}}),~\phi_2=\frac{\Delta \omega t_{BS}}{2}-\frac{\pi}{2}{\rm sgn }(\Delta \omega) ~,
\label{S3}
\end{eqnarray}
where $ T $ and $ R $ are determined by the Eqs. (\ref{8}) of the paper. It can be seen that the coefficients $ T $ and $ R $ satisfy the condition $ T + R = 1 $ and, by sense, are the coefficients of transmission and reflection of photons, respectively. It should be added that the phases $ \phi_1 $ and $ \phi_2 $ are not essential for the description of the effect under consideration. Next, we consider the Eqs. (\ref {S3}) taking into account the time delay $ \tau $ between 1 and 2 photons and the BS offset from the equilibrium position. It is more convenient to count the time $ t_1 $ and $ t_2 $ relative to 1 and 2 detectors, respectively. For this, it is necessary to take into account the distance $ x_1 $ from 1 detector to BS and the distance $ x_2 $ from 2 detectors to BS (of course, considering the exit points from BS, 1 and 2 photons to be the same, see Fig. \ref{fig_1}). As a result, the time $ t_1 $ and $ t_2 $ in the Eq. (\ref {S3}) should be replaced by $ t_1 \to t_1- x_1 / c $ and $ t_1 \to t_2- x_2 / c $, respectively. With the equilibrium position BS $ x_1 = x_2 $, and when it is shifted, the difference $ \delta x = x_2-x_1 = \delta \tau c $ in the distance from BS to 1 and 2 detectors occurs. Therefore, when BS is shifted (for example, upward, see Fig. \ref{fig_1}), the first photon reflected from BS will travel a larger path to detector D2 (which means $ \delta \tau / 2 $ later) than the second photon reflected from BS to detector D1 (means earlier on $ \delta \tau / 2 $). We assume that the first photon is delayed by the time $ \tau $ relative to the second, in this case this photon will be reflected from BS by the time $ \tau $ later. As a result, it is easy to obtain the Eq. (\ref {7}) of an article.

\begin{thebibliography}{150}
\bibitem{HOM_1987} C. K. Hong, Z. Y. Ou and L. Mandel, Measurement of Subpicosecond Time Intervals between Two Photons by Interference. \emph{Phys. Rev. Lett.} {\bf 59} 2044--2046  (1987).
\bibitem{Nielsen_2000} M.  A.  Nielsen  and  I.  L.  Chuang, Quantum Computation and Quantum Information. \emph{Cambridge University Press} (2000).
\bibitem{Gisin_2002} N. Gisin, G. Ribordy, W. Tittel, and H. Zbinden, Quantum cryptography, \emph{Rev. Mod. Phys.} {\bf 74} 145--195 (2002).
\bibitem{Sangouard_2011} N. Sangouard, C. Simon, H. de Riedmatten, and N. Gisin, Quantum repeaters based on atomic ensembles and linear optics. \emph{Rev. Mod. Phys.} {\bf 83} 33--80  (2011). 
\bibitem{Mandel_1995} L. Mandel and E. Wolf, Optical Coherence and Quantum Optics (Cambridge University Press, Cambridge, 1995)
\bibitem{Scully_1997} Scully M.O. and Zubairy, M.S. (1997) Quantum Optics. Cambridge University Press, Cambridge.
\bibitem{Campos_1989} Richard A. Campos, Bahaa E. A. Saleh, and Malvin C. Teich, Quantum-mechanical lossless beam splitter: SU(2) symmetry and photon statistics, \emph{Phys. Rev. A} {\bf 40} 1371 (1989).
\bibitem{Luis_1995} A. Luis and L. L. Sanchez-Soto, A quantum description of the beam splitter, \emph{J. Opt. B: Quantum Semiclass. Opt.} {\bf 7} 153 (1995).
\bibitem{Fearn_1989} H. Fearn and R. Loudon, Theory of two-photon interference, \emph{J. Opt. Soc. Am. B} {\bf 6} 917-927 (1989).
\bibitem{Aephraim_1992} A. M. Steinberg, P. G. Kwiat, and R. Y. Chiao, Dispersion cancellation and high-resolution time measurements in a fourth-order optical interferometer. \emph{Phys. Rev. A} {\bf 45} 6659 (1992).
\bibitem{Legero_2003} T. Legero, T. Wilk, A. Kuhn, G. Rempe, Time-resolved two-photon quantum interference. \emph{Applied Physics B} {\bf 77} 797–802 (2003).
\bibitem{Wang_2006} Kaige Wang, Quantum theory of two-photon wavepacket interference in a beamsplitter. \emph{J. Phys. B: At. Mol. Opt. Phys.} {\bf 39} R293 (2006).
\bibitem{Legero_2004} T. Legero, T. Wilk, M. Hennrich, G. Rempe, and A.Kuhn, Quantum Beat of Two Single Photons. \emph{Phys. Rev. Lett.} {\bf 93} 070503 (2004).
\bibitem{Longo_2010} P. Longo, P. Schmitteckert and Kurt Busch, Few-Photon Transport in Low-Dimensional Systems: Interaction-Induced Radiation Trapping. \emph{Phys. Rev. Lett.} {\bf 104} 023602 (2010).
\bibitem{Makarov_2017_adf} D. N. Makarov, High Intensity Generation of Entangled Photons in a Two-Mode Electromagnetic Field. \emph{Annalen der Physik} {\bf 529} 1600408 (2017).
\bibitem{Makarov_SREP_2018} D. N. Makarov, Quantum entanglement of a harmonic oscillator with an electromagnetic feld. \emph{Scientific Reports} {\bf 8} 8204  (2018).
\bibitem{Makarov_2019_PRA} D. N. Makarov, Optical-mechanical cooling of a charged resonator, \emph{Phys. Rev. A} {\bf 99}(3), 033850 (2019).
\bibitem{Makarov_2018_PRE} D. N. Makarov, Coupled harmonic oscillators and their quantum entanglement. \emph{Phys. Rev. E} {\bf 97} 042203 (2018).
\bibitem{Tey_2008} M. K. Tey, Z. Chen, et al. Strong interaction between light and a single trapped atom without the need for a cavity. \emph{Nature Physics} {\bf 4} 924 - 927 (2008).
\bibitem{Lyons_2018} A. Lyons, G. C. Knee, E. Bolduc et al. Attosecond-resolution Hong-Ou-Mandel interferometry. \emph{Science Advances} {\bf 4} 9416 (2018).
\bibitem{Grice_1997} W. P. Grice and I. A. Walmsley, Spectral information and distinguishability in type-II down-conversion with a broadband pump. \emph{Phys. Rev. A} {\bf 56} 1627 (2000).
\bibitem{Erdmann_2000} R. Erdmann, D. Branning, W. Grice and I. A. Walmsley, Restoring dispersion cancellation for entangled photons produced by ultrashort pulses. \emph{Phys. Rev. A} {\bf 62} 053810 (2000).
\bibitem{Barbieri_2017} M. Barbieri, E. Roccia1, L. Mancino1 et al. What Hong-Ou-Mandel interference says on two-photon frequency entanglement. \emph{Sci. Rep.} {\bf 7} 7247 (2017).
\bibitem{Shih_1999} Y. H. Shih, in Advances in Atomic, Molecular, and Optical Physics, ed., B. Bederson and H. Walther, Academic Press, Cambridge, Vol. 41, 2-42 (1999).
\bibitem{Kwiat_1992} P. G. Kwiat, A. M. Steinberg and R. Y.Chiao, Observation of a ``quantum eraser": A revival of coherence in a two-photon interference experiment. \emph{Phys. Rev. A} {\bf 45} 7729--7739 (2000).
\bibitem{Tambasco1_2018} Jean-Luc Tambasco1, G. Corrielli, R. J. Chapman et al. Quantum interference of topological states of light. \emph{Science Advances} {\bf 4} eaat3187 (2018).
\bibitem{Law_2000} C. K. Law, I. A. Walmsley and J. H. Eberly, Continuous Frequency Entanglement: Effective Finite Hilbert Space and Entropy Control \emph{Phys. Rev. Lett.} {\bf 84} 5304--5307 (2000).
\bibitem{Ekert_2002} A. K. Ekert, C. M. Alves, Daniel K. L. Oi, M. Horodecki, P. Horodecki and L. C. Kwek,  Direct estimations of linear and nonlinear functionals of a quantum state. \emph{Phys. Rev. Lett.} {\bf 88}  217901  (2002).
\bibitem{Lee_2003} J. Lee, M. S. Kim, G. Brukner,  Operationally invariant measure of the distance between quantum states by complementary measurements \emph{Phys. Rev. Lett.} {\bf 91}  087902  (2003).
\end{thebibliography}
\end{document}